# Asymmetric Ground States in La$_{0.67}$Sr$_{0.33}$MnO$_3$/BaTiO$_3$ heterostructures Induced by Flexoelectric Bending

**AUTHORS**

Mingqun Qi,[1,2] Zhen Yang,[2,3] Shengru Chen,[2,3] Shan Lin,[2,3] Qiao Jin,[2,3] Haitao Hong,[2,3] Dongke Rong,[2] Haizhong Guo,[4] Can Wang,[2,3,5] Ziyu Wang,[6] Kui-juan Jin,[2,3,5,a)] Zhenping Wu,[1,a)] and Er-Jia Guo[2,3,5,a)]

**AFFILIATIONS**

[1] State Key Laboratory of Information Photonics and Optical Communications and Laboratory of Optoelectronics Materials and Devices, School of Science, Beijing University of Posts and Telecommunications, Beijing 100876, China

[2] Beijing National Laboratory for Condensed Matter Physics and Institute of Physics, Chinese Academy of Sciences, Beijing 100190, China

[3] University of Chinese Academy of Sciences, Beijing 100049, China

[4] Key Laboratory of Material Physics, Ministry of Education, School of Physics and Microelectronics, Zhengzhou University, Zhengzhou 450001, China

[5] Songshan Lake Materials Laboratory, Dongguan, Guangdong 523808, China

[6] The Institute of Technological Sciences, Wuhan University, Wuhan 430072, P. R. China

[a)] Corresponding author. Emails: kjjin@iphy.ac.cn, zhenpingwu@bupt.edu.cn, and ejguo@iphy.ac.cn

**ABSTRACT**

Misfit strain delivered from single-crystal substrates typically modifies the ground states of transition metal oxides, generating increasing interests in designing modern transducers and sensors. Here, we demonstrate that magnetotransport properties of $La_{0.67}Sr_{0.33}MnO_3$ (LSMO) films were continuously tuned by uniaxial strain produced by a home-designed bending jig. The electrical conductivity and Curie temperature of LSMO films are enhanced by bending stresses. The resistivity of a *u-shape* bended LSMO decays three times faster than that of a *n-shape* bended LSMO as a response to the same magnitude of strain. The asymmetric magnetic states in uniaxially strained LSMO are attributed to the dual actions of Jahn-Teller distortion and strain gradient mediated flexoelectric fields in an adjacent ferroelectric layer. These findings of multi-field regulation in a single material provide a feasible means for developing flexible electronic and spintronic devices.

**TEXT**

Flexible electronics have witnessed impressive progress in wearable sensing, environmental monitoring, energy chemistry, and biomedicine. [1-7] As listed in one of the top ten scientific and technological projects, flexible materials and their related devices inevitably promote a major industrial revolution in the future. [8] By then, our daily life and work will be greatly benefited from these innovation technologies. [9] Typically, electronic devices mainly make use of their single physical properties of existing flexible materials, such as high conductivity, high transmission, and *etc*. [10-12] The integration of complex materials with multiple properties are essential for developing smaller, smarter, and low-power consuming devices. Up to now, the flexible devices based on the complex oxides have not been explored extensively due to the limitation of their rigid and single-crystalline substrates. [13] Typically, these substrates are used for oxide thin film growth due to the compatible crystallographic symmetry and small lattice mismatch. However, these single crystal substrates are extremely fragile and non-foldable. The as-grown thin films are inflexible even though the thickness of substrates was reduced to sub-micrometer scale. This fact inevitably limits the investigations of wide-range strain effects on their physical properties and prevents their potential applications towards flexible devices.

Van der Waals (vdW) heteroepitaxy has been proposed initially to demonstrate the epitaxial growth of two-dimensional (2D) layered materials on other 2D layered materials. This method has been extended to describe the epitaxial growth of functional oxides on 2D materials. [14-16] For instance, Chu's group had successfully fabricated various oxides on the layered fluoropolytic-mica (Mica) substrates. [17-20] One of the key advantages of vdW epitaxy is the weak interaction between films and substrates, leading to a large tolerance in the lattice mismatch and crystallographic symmetry. [21] Therefore, oxides with different crystallographic structures can be epitaxially grown on Mica with relatively high quality. [22-25] These oxide films are no longer clamped by Mica substrates, resulting in a nearly strain-free virgin state. Most importantly, the layered character of Mica makes it very flexible and easy to be folded under large bending stress, providing a playground for studying strain effects on the physical properties of transition metal oxides. The high thermal and chemical stability also allows Mica to be used as appropriate high-temperature substrates for thin-film growth. [21] In this Letter, we fabricated the heterostructures composed of ferromagnetic $La_{0.67}Sr_{0.33}MnO_3$ (LSMO) and ferroelectric $BaTiO_3$ (BTO) layers. The magnetoelectric characteristics of LSMO were investigated by applying uniaxial strain ranging from -2 to 2% using a home-designed bending jig. We find that the conductivity and Curie temperature ($T_C$) exhibit an asymmetric response under the same magnitude of bending stress. The unique behavior is attributed to both intrinsic structural distortions induced electronic modification and extrinsic ferroelastic effects from an adjacent ferroelectric layer.

The heterostructures composed of BTO and LSMO were grown on (001)-oriented Mica and (111)-oriented STO substrates simultaneously using pulsed laser deposition (PLD). [26] The LSMO/BTO and BTO/LSMO/BTO trilayers were fabricated for strain-tuning. The thicknesses of BTO and LSMO thin layers are 15 and 50 nm, respectively. The BTO layer was used as a buffer layer to minimize misfit strain between LSMO and Mica. [27] During the deposition, the substrate's temperature was kept at 750 °C and oxygen partial pressure was maintained at 200 mTorr. The laser frequency and energy density were kept as 5 Hz and ~1.0 J/cm$^2$, respectively. The samples were cooled down slowly to room temperature at a rate of -5 °C/min under the oxygen partial pressure of 100 Torr. X-ray diffraction (XRD) measurements were performed using a Bruker D8 Discovery high-resolution diffractometer with Cu Kα1 radiation (Figure S1, Supplemental Materials). XRD $\theta$–$2\theta$ scans reveal that all layers are epitaxially grown without impurity phases. Reciprocal space mapping around Mica substrate's 00$l$ reflections was measured using a Panalytical X'Pert MRD four-circle X-ray diffractometer. RSM results demonstrate that LSMO/BTO bilayers are epitaxially grown along the (111) orientation. We calculate the inter-planar spacings for LSMO (111) and BTO (111) layers on (111)-STO substrates are 2.224 Å (~2.254 Å for LSMO bulk), 2.281 Å (~2.314 Å for BTO bulk), respectively. The LSMO film is under the substrate-induced tensile-strain, leading to the out-of-plane lattice constant shrinks by ~ 1.3%. In contrast, the BTO and LSMO films grown on Mica is almost strain-free, similar to the previous reports. Therefore, the virgin strain state of as-grown LSMO films on Mica substrates is close to its bulk form. [28]

The electrical transport and magnetic properties of LSMO films on (111)-STO and Mica substrates were measured using a physical properties measurement system (PPMS) and magnetic property measurement system (MPMS), respectively. Figure 1(a) shows the *M-T* curves of LSMO/BTO bilayers at an in-plane magnetic field of 1 kOe. The $T_C$ of LSMO films on Mica is ~ 350 K, whereas the $T_C$ of LSMO films on (111)-STO is ~ 370 K, which is close to the $T_C$ of bulk LSMO. [29] The lower $T_C$ of LSMO films on Mica is likely due to the different microstructures and strain states in the as-grown films. Room-temperature magnetic hysteresis loops yield the coercive fields of both LSMO/BTO bilayers are around 100 Oe (inset of Figure 1b). The LSMO films grown on (111)-STO shows a higher saturation moment than those of LSMO films grown on Mica. The resistivities ($\rho$) and magnetoresistances [MR = ($\rho_H-\rho_0$)/$\rho_0$] were measured using the van der Pauw method, where $\rho_0$ and $\rho_H$ are the resistivities under magnetic fields of 0 and 9 T, respectively. The resistivities of LSMO films obtained from two geometries, i. e. one is the current passed along the bending direction and another is the current passed perpendicular to the bending direction, are almost equivalent. Thus, the resistivities of LSMO films were averaged from those values. Figure 1(c) shows the $\rho-T$ curves of both LSMO/BTO bilayers. The LSMO layers exhibit clear insulator-to-metal transition as decreasing temperature. This is a typical character of the hole-doped manganites. [30-32] The LSMO films on Mica exhibit a larger $\rho$ than that of LSMO on (111)-STO substrates. At 10 K, ρ of LSMO on Mica is more than one order of magnitude larger than that of LSMO on STO. Field-dependent MR at 10 and 300 K were present in Figures 1(d). The MR of LSMO films shows negative values under external fields, demonstrating the ferromagnetic characteristic of LSMO. At 10 K, the MR shows symmetric peaks at small fields corresponding to its coercive fields ($H_C$). When $T$ increases to 300 K, the MR of LSMO on Mica reduces slightly and the butterfly-like hysteresis loop disappears. Instead, the MR keeps nearly unchanged at small fields and starts to decrease linearly when the magnetic fields exceed $H_C$.

Typically, Mica can be mechanically exfoliated down to 10 µm in thickness due to the interlamellar vdW bonding character. [21, 33] The ultrathin Mica allows them to be bended substantially. We designed a mechanical bending jig to apply uniaxial strain to the samples. The strain states of as-grown films can be *in-situ* controlled and maintained by bending forces during the measurements (Figure S2, Supplemental Materials). The center position of samples was bended in "*u*-shape" or "*n*-shape". A high-resolution camera with a fixed position was used to capture the sample's positions and bending curvatures under different strain states (Figure S3, Supplemental Materials). [26, 34] We quantified the uniaxial strain of LSMO films induced by mechanical forces using bending parameters, e. g. the length of sample's bending part and the deformation displacement at the center positions. The strain states are directly linked to their physical properties. Figures 2a and 2b show the zero-field $\rho-T$ curves of LSMO films under switchable bending stresses, respectively. The maximum bending stresses were achieved up to −2% and 1.8%, respectively. The $\rho$ of LSMO films reduces progressively by increasing uniaxial stress. Figure 2c summarizes the change rate of resistivity [$\Delta\rho = (\rho_{strain}-\rho_{virgin})/\rho_{virgin}$] as a function of stress at 150, 300 and 350 K. Compared to the tensile strain cases, the $\rho$ of LSMO films decays rapidly when the uniaxial stress reduces from virgin state to −1% and slows down when compressive strain beyond −1%. The $\Delta\rho(\varepsilon)$ in compressive stress regime is almost three times larger than that of tensile stress regime. $T_C$ enhances at nearly a constant ratio of ~ 5 K per 1% as increasing the bending stress (Figure 2d).

Intriguing magnetotransport properties of LSMO films were further investigated when LSMO films were under different stresses. Figure 3a shows the typical butterfly-like hysteresis loops of MR at 10 K when the stress switches from the compressive state ($\varepsilon = -2\%$) to the tensile state ($\varepsilon = 1.8\%$). MR measured at different stress states exhibits similar variation trends, i. e. positive MR at small fields and negative MR at high fields. We derive $H_C$ from the symmetric peaks of MR. Figure 3b presents the $H_C$ of LSMO films as a function of stress at various temperatures. With increasing temperature and stress, $H_C$ decreases distinctly. The $H_C$ of LSMO films under compressive stresses generally is smaller than that of the tensile-stressed LSMO films. In particular, we observe unique double symmetric peaks on MR loops when $\varepsilon = -2\%$. [36, 37] We hypothesize that the LSMO films break into two magnetic layers that exhibit different $H_C$ when the uniaxial strain exceeds 2%. Figure 3c summarizes the MR as a function of stress at 10 K when $\mu_0 H = 0.5$ and 3 T. The stress reduces MR($\mu_0 H = 3$ T) by a factor of two and this effect decreases at small magnetic fields.

The stress dependent asymmetric behaviors in magnetotransport properties of LSMO films demonstrate that additional impact factors that control the ground states of LSMO films. The basic unit cell of Mica contains two silicate tetrahedral sheets ($SiO_4$) on both sides of an aluminum octahedron ($AlO_6$). The layers with strong covalent bonding, however, are weakly stacked together by interlayer cations, resulting in a vdW gap of ~ 1 nm. The cleavage of Mica along the vdW gap layer produces the fresh, atomically flat surface with randomly distributed cations. These cleaved Mica substrates preserve the charge neutral surfaces which are ideal for vdW oxide epitaxy. Experimentally, we determine that the BTO films grown on Mica exhibit an intrinsic upwards ferroelectric polarization using a piezoelectric force microscope (Asylum Research MFP3D) (Figure S4, Supplemental Materials). Mechanically bending forces will generate a strain gradient along the in-plane direction, resulting in a flexoelectric field perpendicular to the surface plane. The magnitude of the flexoelectric field is proportional to the strain gradient, i. e.

the bending curvature. [38-40] Apparently, the "*u*-shape" and "*n*-shape" bending statuses produce opposite flexoelectric fields, which are responsible for the asymmetric ground states of LSMO under compressive and tensile strains. The flexoelectric modulation to transport properties of LSMO is similar to the ferroelectric control of conductivity at BTO/LSMO interface. In the "*u*-shape" bending status, the flexoelectric field in BTO layers points upward, thus the electrons accumulate at the LSMO/BTO interfaces. Thus, the $\rho$ reduces rapidly and $T_C$ enhances dramatically in the compressively stressed LSMO films compared to those of tensile-stressed LSMO cases (corresponding to the hole accumulation). However, although the flexoelectric field effects play a significant role in the small strain regime, we believe the modulation of $\rho$ should be dominated by uniaxial strain-induced Jahn-Teller distortion instead of the flexoelectric polarization in a large strain regime. Therefore, the reduction of $\rho$ is almost the same under both maximum bending stresses.

To eliminate the flexoelectric effects on the transport properties of LSMO films, we sandwiched a LSMO film between two BTO layers with equal thickness. The top and bottom BTO layers generate identical flexoelectric polarization within the layers. Therefore, one interface of LSMO layers is hole accumulation, while the other interface is hole depletion, or vice versa, depending on the bending status. The magnitude of interfacial charge doping effects is equal when LSMO films in both bending statuses. Figure 4a shows the $\rho-T$ curves of BTO/LSMO/BTO trilayers at virgin state, compressive strain state, and tensile strain state. The virgin LSMO film presents an insulator-to-metal transition at a temperature of ~ 347 K, which is identical to that of LSMO/BTO bilayers. The $\rho$ of a BTO/LSMO/BTO trilayer is three times larger than the resistivity of LSMO/BTO bilayers and meanwhile, it exhibits a low-temperature upturn. These results may attribute to the structural distortion induced weak localization. [41, 42] We notice that the $\rho$ of BTO/LSMO/BTO trilayers reduces only 10% and the $T_C$ increases only ~ 1 K upon the mechanical bendings, in sharp contrast to the results of LSMO/BTO bilayers, further highlight the previously discussed flexoelectric effects at small stress regime. The genuine uniaxial strain effects on magnetotransport properties of LSMO films were further investigated under different strain states (Figure S5, Supplemental Materials). Similar to the bilayer's results, $H_C$ reduces as increasing uniaxial stress. Figure 4b presents MR as a function of magnetic fields when ε = −1%, 0, and 1%. The symmetric peaks at small fields appear in the hysteresis loop when the trilayer is in the virgin and compressive-stress states. However, we notice that the MR curves of tensile-strained LSMO films exhibit four distinct peaks. These results suggest that the LSMO films possibly break into two magnetic layers under large uniaxial tensile strains. Different parts of LSMO films exhibit independent magnetic switching behaviors. Thus, the MR curves possess multiple peaks, corresponding to different $H_C$ in the hysteresis loops.

In summary, we report the dual actions of uniaxial stress and flexoelectric fields on the transport properties of LSMO films in proximity to ferroelectric materials. The uniaxial stresses were *in-situ* continuously controlled in a wide range from −2 to 1.8% using a home-designed bending jig. We find that the stress effect on the electrical conductivity and magnetoresistance is nonlinear. We hypothesize that the asymmetric magnetoelectric responses can be attributed to the flexoelectric fields generated by uniaxial stress, which manifests the carrier density of LSMO films. Our results shed intensive insights on the strong competition between charge and lattice degrees of freedom in the functional oxides. Additionally, the design of highly stress-sensitive flexible devices paves a potential route toward the neoteric strain sensors.

See the supplementary material for additional information about the structural characterizations, bending jig photograph, calibration of bending stress, piezoresponse force microscopy characterizations, magnetoresistance measurements.

**ACKNOWLEDGEMENTS**

This work was supported by the National Key Basic Research Program of China (Grant

Nos. 2020YFA0309100 and 2019YFA0308500), the National Natural Science Foundation of China (Grant Nos. 12074044, 11974390, and 11721404), the Excellent Youth Foundation of Hubei Province (Grant Number: 2019CFA083), the Beijing Nova Program of Science and Technology (Grant No. Z191100001119112), the Beijing Natural Science Foundation (Grant No. 2202060), the Guangdong-Hong Kong-Macao Joint Laboratory for Neutron Scattering Science and Technology, the Fund of State Key Laboratory of Information Photonics and Optical Communications (IPOC2021ZT05), and the Strategic Priority Research Program (B) of the Chinese Academy of Sciences (Grant No. XDB33030200).## AUTHOR DECLARATIONS

**Conflict of Interest**

The authors declare no competing financial interest.

**Data Availability**

The data that support the findings of this study are available from the corresponding authors upon reasonable request.

## REFERENCES

[1] B. D. Gates, Science 323, 1566-1567 (2009).

[2] A. Nathan, A. Ahnood, M. T. Cole, S. Lee, Y. Suzuki, P. Hiralal, F. Bonaccorso, T. Hasan, L. Garcia-Gancedo, A. Dyadyusha, S. Haque, P. Andrew, S. Hofmann, J. Moultrie, D. P. Chu, A. J. Flewitt, A. C. Ferrari, M. J. Kelly, J. Robertson, G. A. J. Amaratunga, and W. I. Milne, Proc. IEEE 100, 1486-1517 (2012).

[3] H. B. Li, Y. J. Ma, and Y. G. Huang, Mater. Horiz. 8, 383-400 (2021).

[4] W. Gao, H. Ota, D. Kiriya, K. Takei, and A. Javey, Acc. Chem. Res. 52, 523-533 (2019).

[5] K. X. Teng, Q. An, Y. Chen, Y. H. Zhang, and Y. T. Zhao, Acc. Chem. Res. 1302-1337 (2021).10

[6] X. D. Chen, J. A. Rogers, S. P. Lacour, W. P. Hu, and D. H. Kim, Chem. Soc. Rev. 48, 1431-1433 (2019).

[7] S. Choi, H. Lee, R. Ghaffari, T. Hyeon, and D. H. Kim, Adv. Mater. 28 , 4203-4218 (2016).

[8] J. He and T. M. Tritt, Science 357, caak9997 (2017).

[9] S. Park, G. Wang, B. Cho, Y. Kim, S. Song, Y. Ji, M. H. Yoon, and T. Lee, Nat. Nanotechnol. 7, 438-442 (2012).

[10] E. B. Secor, P. L. Prabhumirashi, K. Puntambekar, M. L. Geier, and M. C. Hersam, J. Phys. Chem. Lett. 4, 1347-1351 (2013).

[11] Y. Cui, Z. H. Qin, H. Wu, M. Li, and Y. J. Hu, Nat. Commun. 12, 1284 (2021).

[12] S. Macher, M. Schott, M. Dontigny, A. Guerfi, K. Zaghib, U. Posset, and P. Lobmann, Adv. Mater. Technol. 6, 2000836 (2021).

[13] T. G. Sano, T. Yamaguchi, and H. Wada, Phys. Rev. Lett. 118, 178001 (2017).

[14] H. Kum, D. Lee, W. Kong, H. Kim, Y. Park, Y. Kim, Y. Baek, S. H. Bae, K. Lee, and J. Kim, Nat. Electron. 2, 439-450 (2019).

[15] S. H. Bae, H. Kum, W. Kong, Y. Kim, C. Choi, B. Lee, P. Lin, Y. Park, and J. Kim, Nat. Mater. 18, 550-560 (2019).

[16] D. C. Geng and H. Y. Yang, Adv. Mater. 30, 1800865 (2018).

[17] V. Q. Le, T. H. Do, J. R. D. Retamal, P. W. Shao, Y. H. Lai, W. W. Wu, J. H. He, Y. L. Chueh, and Y. H. Chu, Nano Energy 56, 322-329 (2019).

[18] B. R. Tak, V. Gupta, A. K. Kapoor, Y. H. Chu, and R. Singh, ACS Appl. Electron. Mater. 1, 2463-2470 (2019).

[19] M. Yen, Y. H. Lai, C. Y. Kuo, C. T. Chen, C. F. Chang, and Y. H. Chu, Adv. Funct. Mater. 30, 2004597 (2020).

[20] T. D. Ha, J. W. Chen, M. Yen, Y. H. Lai, B. Y. Wang, Y. Y. Chin, W. B. Wu, H. J. Lin, J. Y. Juang, and Y. H. Chu, ACS Appl. Mater. Interfaces 12, 46874-46882 (2020).

[21] Y. F. He, L. X. Wang, Z. X. Xiao, Y. W. Lv, L. Liao, and C. Z. Jiang, Chin. Phys.

Lett. 37, 088502 (2020).

[22]  H. Xu, Z. L. Luo, C. G. Zeng, and C. Gao,  Chin. Phys. Lett. 36, 078101 (2019).

[23]  R. Q. Cheng, Y. Wen, L. Yin, F. M. Wang, F. Wang, K. L. Liu, T. A. Shifa, J. Li, C. Jiang, Z. X. Wang, and J. He,  Adv. Mater. 29, 1703122 (2017).

[24]  T. Amrillah, Y. Bitla, K. Shin, T. N. Yang, Y. H. Hsieh, Y. Y. Chiou, H. J. Liu, T. H. Do, D. Su, Y. C. Chen, S. U. Jen, L. Q. Chen, K. H. Kim, J. Y. Juang, and Y. H. Chu,  Acs Nano 11, 6122-6130 (2017).

[25]  Y. T. Zhang, Y. Q. Cao, Hh Hu, X. Wang, P. Z. Li, Y. Yang, J. Zheng, C. Zhang, Z. Q. Song, A. D. Li, and Z. Wen,  ACS Appl. Mater. Interfaces 11, 8284-8290 (2019).

[26]  C.Cheng, S.R.Chen, J.Deng, G.Li, Q.H.Zhang, G.Lin, T.P.Ying, E.J.Guo, J.G.Guo, X.L.Chen,  Chin. Phys. Lett. 39, 047301 (2022)

[27]  M. K. Lee, T. K. Nath, C. B. Eom, M. C. Smoak, and F. Tsui,  Appl. Phys. Lett. 77, 3547-3549 (2000).

[28]  H. Baaziz, N. K. Maaloul, A. Tozri, H. Rahmouni, S. Mizouri, K. Khirouni, and E. Dhahri,  Chem. Phys. Lett. 640, 77-81 (2015).

[29]  P. Graziosi, A. Gambardella, M. Prezioso, A. Riminucci, I. Bergenti, N. Homonnay, G. Schmidt, D. Pullini, and D. Busquets-Mataix,  Phys. Rev. B 89, 214411 (2014).

[30]  M. Fath, S. Freisem, A. A. Menovsky, Y. Tomioka, J. Aarts, and J. A. Mydosh,  Science 285, 1540-1542 (1999).

[31]  X. Hong, A. Posadas, A. Lin, and C. H. Ahn,  Phys. Rev. B 68, 134415 (2003).

[32]  I. Stolichnov, S. W. E. Riester, E. Mikheev, N. Setter, A. W. Rushforth, K. W. Edmonds, R. P. Campion, C. T. Foxon, B. L. Gallagher, T. Jungwirth, and H. J. Trodahl,  Phys. Rev. B 83, 115203 (2011).

[33]  C. Zhang, S. S. Ding, K. M. Qiao, J. Li, Z. Li, Z. Yin, J. R. Sun, J. Wang, T. Y. Zhao, F. X. Hu, and B. G. Shen,  ACS Appl. Mater. Interfaces 13, 28442-28450 (2021).

[34]  X. Wang, A. Y. Cui, F. F. Chen, L. P. Xu, Z. G. Hu, K. Jiang, L. Y. Shang, and J. H. Chu,  Small 15, 1903106 (2019).

[35] S. S. Hong, M. Q. Gu, M. Verma, V. Harbola, B. Y. Wang, D. Lu, A. Vailionis, Y. Hikita, R. Pentcheva, J. M. Rondinelli, and H. Y. Hwang, Science 368, 71-76 (2020).

[36] C. Boskovic, M. Pink, J. C. Huffman, D. N. Hendrickson, and G. Christou, J. Am. Chem. Soc. 123, 9914-9915 (2001).

[37] H. Sawada, Y. Morikawa, N. Hamada, and K. Terakura, J. Am. Chem. Soc. 177, 879-880 (1998).

[38] M. M. Yang, D. J. Kim, and M. Alexe, Science 360, 904-907 (2018).

[39] R. Guo, L. You, W. N. Lin, A. Abdelsamie, X. Y. Shu, G. W. Zhou, S. H. Chen, L. Liu, X. B. Yan, J. L. Wang, and J. S. Chen, Nat. Commun. 11, 2571 (2020).

[40] F. Zhang, P. Lv, Y. T. Zhang, S. J. Huang, C. M. Wong, H. M. Yau, X. X. Chen, Z. Wen, X. N. Jiang, C. G. Zeng, J. W. Hong, and J. Y. Dai, Phys. Rev. Lett. 122, 257601 (2019).

[41] Y. Z. Gao, J. C. Zhang, G. X. Cao, X. F. Mi, and H. U. Habermeier, Solid State Commun. 154, 46-50 (2013).

[42] Z. L. Liao, F. M. Li, P. Gao, L. Li, J. D. Guo, X. Q. Pan, R. Jin, E. W. Plummer, and J. D. Zhang, Phys. Rev. B 92, 125123 (2015).

**FIGURES AND FIGURE CAPTIONS**

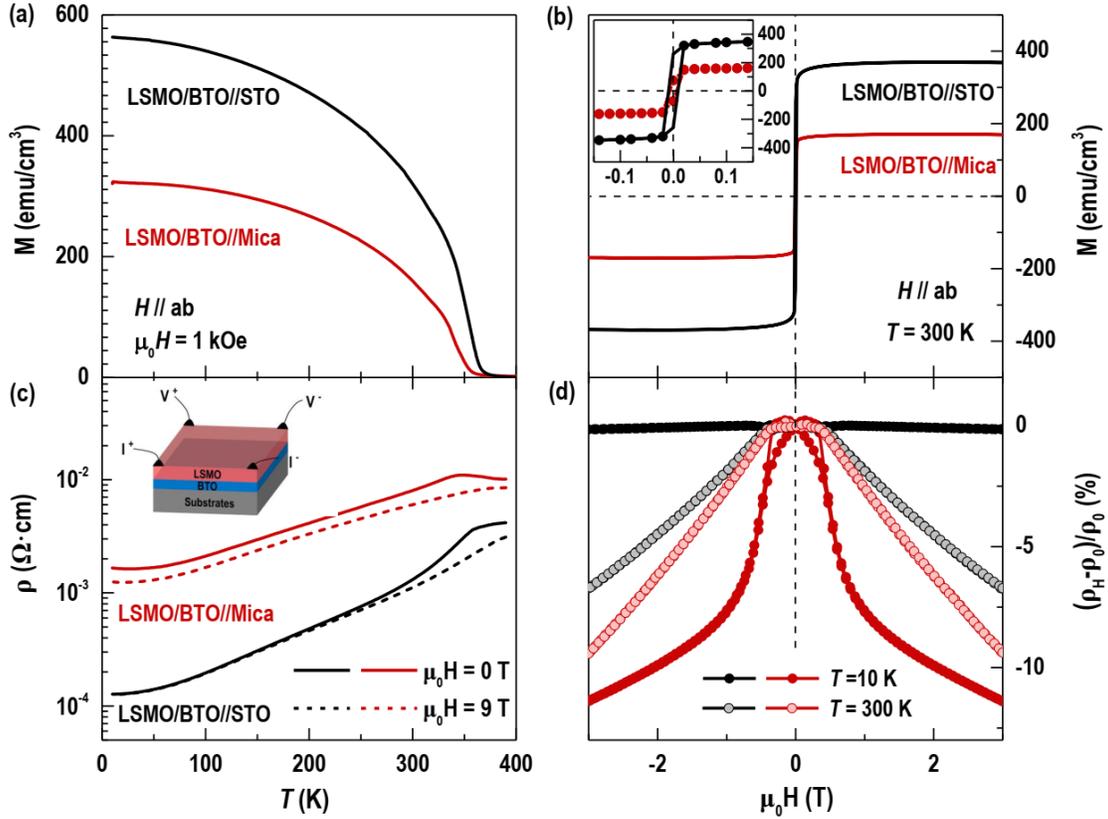

**Figure 1. Transport properties of LSMO/BTO bilayers at virgin states.** Temperature dependent (a) in-plane magnetization (*M*) and (c) resistivities (*ρ*) of LSMO/BTO bilayers on STO (black lines) and Mica (red lines) substrates. *M-T* curves were recorded at 1 kOe after field-cooling. *ρ−T* curves were measured at 0 T(solid line) and 9 T (dashed line), respectively. Inset shows the geometry of electrical measurements using van der Pauw method. Field dependent (b) *M* (300 K) and (d) magnetoresistances [MR = ($\rho_H − \rho_0$) /$\rho_0$] for LSMO/BTO bilayers on STO and Mica substrates at 10 and 300 K.

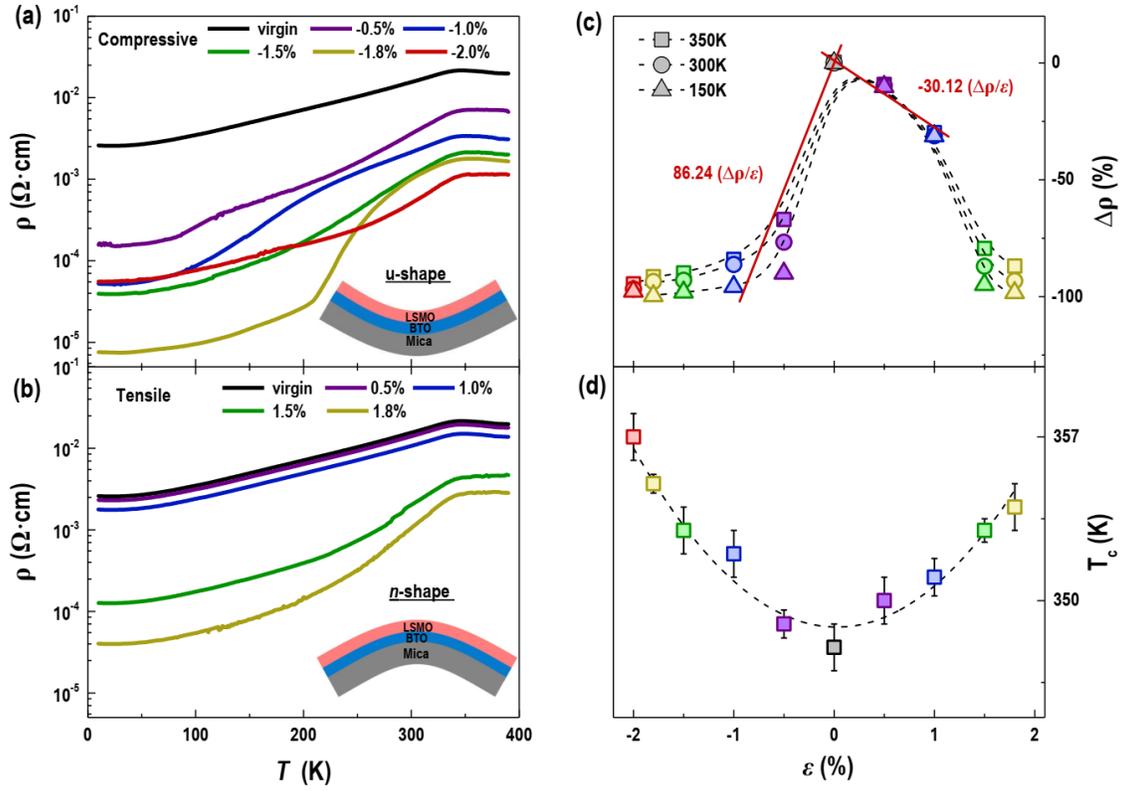

**Figure 2. Tunable ground states of LSMO/BTO bilayers under various uniaxial stress.** Zero-field $\rho$-$T$ curves of LSMO/BTO bilayers under various (a) compressive and (b) tensile stresses. Insets show the schematic of sample status, i. e. *u*-shape and *n*-shape bendings correspond to compressive and tensile stresses, respectively. (c) The resistivity ratios [$\Delta\rho = (\rho_{strain} - \rho_{virgin})/\rho_{virgin}$] and (d) Curie temperatures ($T_C$) were plotted as a function of stress. The $\Delta\rho$ at 150, 300, and 350 K exhibit the similar asymmetric stress dependency.

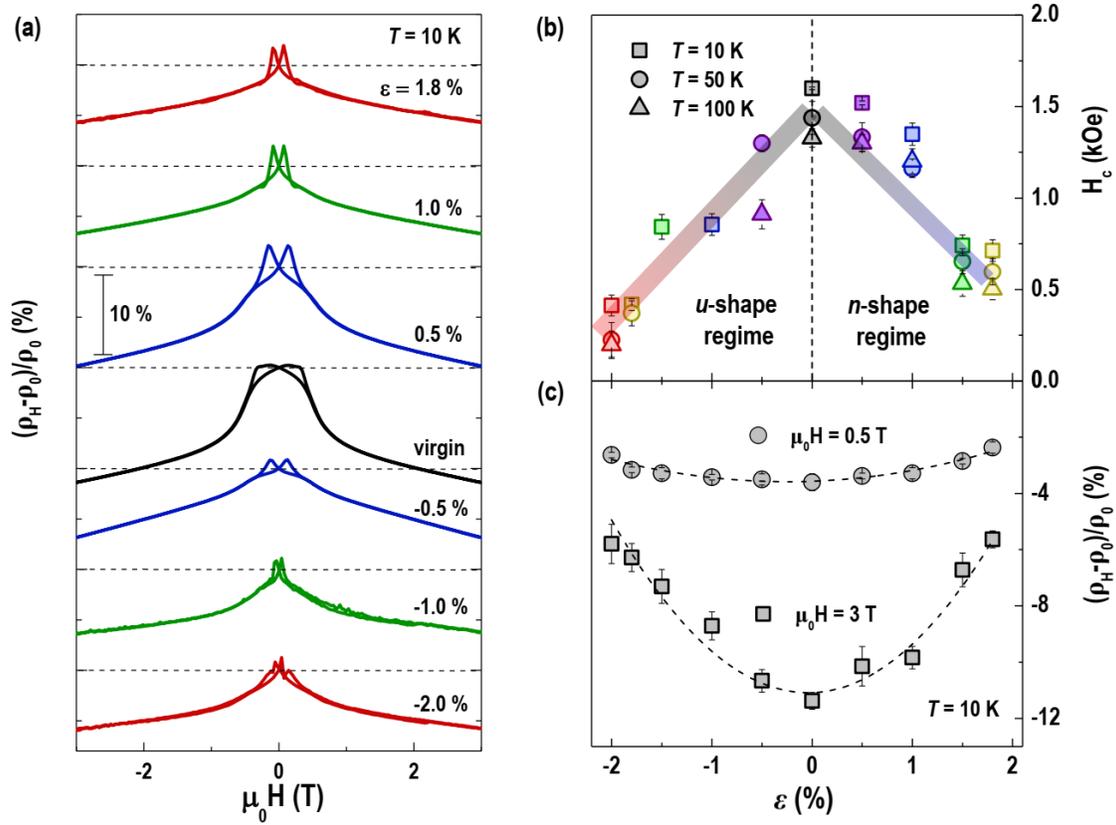

**Figure 3. Magnetotransport properties of a LSMO/BTO bilayer under various stresses**. (a) Field dependent MR at 10 K for the uniaxial stress changes from -2 to 1.8 %. The MR is up-shifted for clarification. Stress dependent (b) coercive fields ($H_C$) and (c) MR. $H_C$ were derived from hysteresis loops recorded at 10, 50, 100, and 150 K. MR were recorded at 10 K under magnetic fields of 0.5 and 3 T.

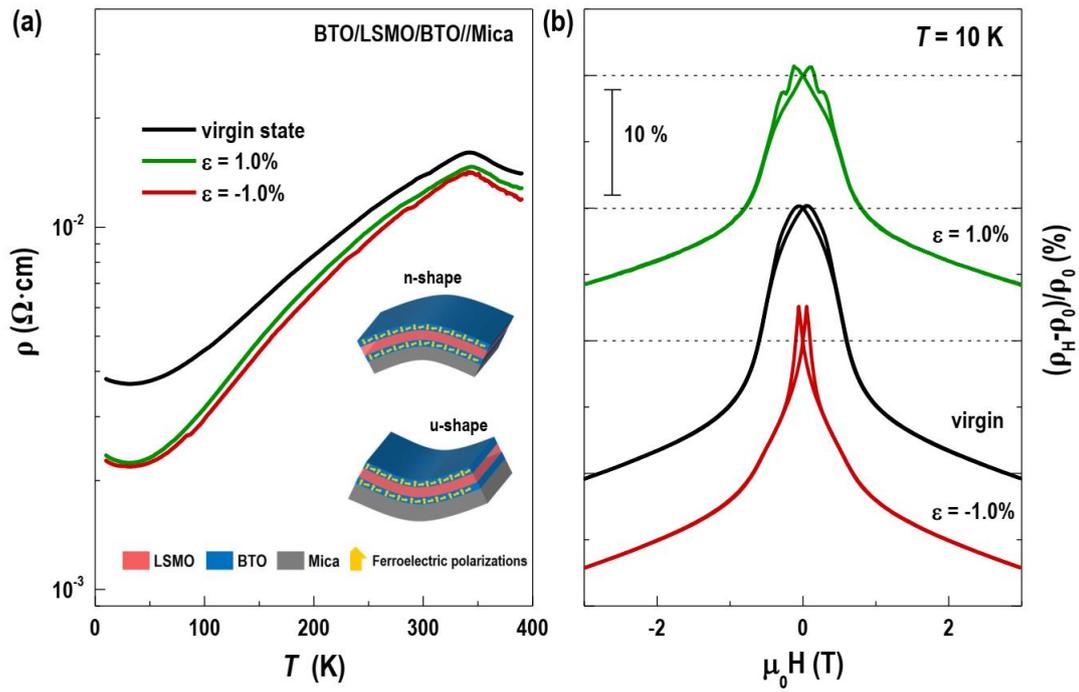

**Figure 4. Magnetotransport measurements on a BTO/LSMO/BTO trilayer under symmetric compression and extension.** (a) Temperature dependent $\rho$ of BTO/LSMO/BTO trilayer under different uniaxial stresses of -1, 0, and 1%. Inset shows the schematics of sample geometry under compression (*u*-shape) and extension (*n*-shape). The yellow arrows indicate the ferroelectric polarization induced by stress-induced flexoelectric fields. (b) Field dependent MR at 10 K under different stresses of -1, 0, and 1%. The MR is up-shifted for clarification.